\documentclass[eng]{csam}
\usepackage{amsmath,amsfonts,amssymb,array,tabularx,enumerate}
\usepackage{mathptmx,amsmath,amssymb,bm}
\usepackage{epsfig, latexsym, graphicx, multirow}
\usepackage{algorithm}
\usepackage{algpseudocode}
\usepackage{algcompatible}

  \usepackage[
    backend=biber,
    style=bwl-FU,
  ]{biblatex}

\submit{draft}

\begin{document}

\title{Improved LARS algorithm for adaptive LASSO in the linear regression model}

\author{Manickavasagar Kayanan\footnote{Corresponding author: Department of Physical Science, University of Vavuniya, Vavuniya, Sri Lanka. E-mail: kayanan@vau.ac.lk}\address[a]{Department of Physical Science, University of Vavuniya, Vavuniya, Sri Lanka;}, Pushpakanthie Wijekoon\address[b]{National Institute of Fundamental Studies, Kandy, Sri Lanka}}

\begin{abstract}
The adaptive LASSO has been used for consistent variable selection in place of LASSO in the linear regression model. In this article, we propose a modified LARS algorithm to combine adaptive LASSO with some biased estimators, namely the Almost Unbiased Ridge Estimator (AURE), Liu Estimator (LE), Almost Unbiased Liu Estimator (AULE), Principal Component Regression Estimator (PCRE), r-k class estimator, and r-d class estimator. Furthermore, we examine the performance of the proposed algorithm using a Monte Carlo simulation study and real-world examples.
\end{abstract}


\section{Introduction}
\label{intro}
Let us consider a linear regression model 
\begin{equation}
  \bm{ y= X\beta+\epsilon}.
\end{equation}
Here, $\bm y$ represents the $n \times 1$ vector of observations on the dependent variable, $\bm X$ is the $n \times p$ matrix of observations on the non-stochastic predictor variables, $\bm \beta$ stands for a $p \times 1$ vector of unknown coefficients, and $\bm \epsilon$ denotes the $n \times 1$ vector of random error terms. These errors are assumed to be independent and identically normally distributed with mean zero and common variance $\sigma^2$.

It is widely acknowledged that the Ordinary Least Squares Estimator (OLSE) serves as the Best Linear Unbiased Estimator (BLUE) for determining the unknown parameter vector in model (1.1), expressed as:
\begin{equation}\begin{split}
    \hat{\bm \beta}_{OLSE}&=\arg\min_{\bm \beta}\left\{(\bm{y-X\beta})'(\bm{y-X\beta})\right\} \\
    &=(\bm X'\bm X)^{-1}\bm X'\bm y.
\end{split}\end{equation}

Nevertheless, the OLSE demonstrates instability and yields parameter estimates with high variance in the presence of multicollinearity within $\bm X$. To mitigate this multicollinearity issue, many researchers resort to biased estimators.

As per Kayanan and Wijekoon (2017), the generalized representation of biased estimators including Ridge Estimator (RE), Almost Unbiased RidgeEstimator (AURE), Liu Estimator (LE), Almost Unbiased Liu Estimator (AULE), Principal Component Regression Estimator (PCRE), r-k class estimator and r-d class estimator can be expressed as:

\begin{equation}\label{gef}
\hat{\bm \beta}_G=\bm G \hat{\bm \beta}_{OLSE}	
\end{equation}where
\[\hat{\bm \beta}_G=\begin{cases}
   \hat{\bm \beta}_{RE}      & \quad \text{if } \bm G=(\bm  X'\bm X+k\bm I)^{-1} \bm  X'\bm X  \\
    \hat{\bm \beta}_{AURE}&\quad \text{if } \bm G=\left(\bm I-k^{2} (\bm  X'\bm X+k\bm I)^{-2} \right)  \\
    \hat{\bm \beta}_{LE}&\quad \text{if } \bm G=(\bm  X'\bm X+\bm I)^{-1} (\bm  X'\bm X+d\bm I)\\
    \hat{\bm \beta}_{AULE}&\quad \text{if } \bm G=\left(\bm I-(1-d)^{2} (\bm  X'\bm X+\bm I)^{-2} \right)  \\
    \hat{\bm \beta}_{PCRE}&\quad \text{if } \bm G=\bm T_{h} \bm T_{h}'\\
    \hat{\bm \beta}_{rk}&\quad \text{if } \bm G=\bm T_{h}\bm  T_{h}' (\bm  X'\bm X+k\bm I)^{-1} \bm  X'\bm X\\
    \hat{\bm \beta}_{rd}&\quad \text{if } \bm G=\bm T_{h} \bm T_{h}'(\bm  X'\bm X+\bm I)^{-1} (\bm  X'\bm X+d\bm I)
  \end{cases}\]

Kayanan and Wijekoon (2017) demonstrated that the r-k class estimator and r-d class estimator outperform other estimators within a specified range of regularization parameter values when multicollinearity exists among the predictor variables. However, biased estimators can introduce substantial bias when the number of predictor variables is high, potentially leading to the inclusion of irrelevant predictor variables in the final model. To address this issue, Tibshirani (1996) proposed the Least Absolute Shrinkage and Selection Operator (LASSO) as

\begin{equation}
    \hat{\bm \beta}_{LASSO}=\arg\min_{\bm \beta}\left\{(\bm{y-X\beta})'(\bm{y-X\beta})\right\} \text{  subject to }\sum_{j=1}^p| \beta_j| \leq t, 
\end{equation}
where $t\geq0$ is a turning parameter. The LASSO solutions has been obtained by the Least Angle Regression (LARS) algorithm.  

According to Zou and Hastie (2005), LASSO failed to outperform Ridge Estimator if high multicollinearity exists among predictors, and it is unsteady when the number of predictors is higher than the number of observations. To overcome this problem, Zou and Hastie (2005) proposed Elastic Net (ENet) estimator by combining  LASSO and RE as  
\begin{equation}
    \hat{\bm \beta}_{Enet}=\arg\min_{\bm \beta}\left\{(\bm{y-X\beta})'(\bm{y-X\beta})+k\sum_{j=1}^p \beta^2_j\right\} \text{  subject to }\sum_{j=1}^p| \beta_j| \leq t. 
\end{equation}The LARS-EN algorithm, which is a modified version of the LARS-LASSO algorithm, has been used to obtain solutions for ENet. 

Further, Zou and Hastie (2005) noted that LASSO does not care about variable importance when a group of variables among which the pairwise correlations are very high.

To handle this problem, Zou (2006) proposed adaptive LASSO by giving different weights to regression coefficients in L1 penalty of LASSO. By taking weight vector $\hat{\bm w}=|\hat{\bm \beta}_{OLSE}|^{-\alpha}$ for any $\alpha>0$, the adaptive LASSO is defined as  
\begin{equation}
    \hat{\bm \beta}_{adpLASSO}=\arg\min_{\bm \beta}\left\{(\bm{y-X\beta})'(\bm{y-X\beta})\right\} \text{  subject to }\sum_{j=1}^p| w_j\beta_j| \leq t. 
\end{equation}
In addition to that Zou and Zhang (2009)  proposed adaptive Enet estimator by combining adaptive LASSO and RE, and it is defined as
\begin{equation}
    \hat{\bm \beta}_{adpEnet}=\arg\min_{\bm \beta}\left\{(\bm{y-X\beta})'(\bm{y-X\beta})+k\sum_{j=1}^p \beta^2_j\right\} \text{  subject to }\sum_{j=1}^p| w_j\beta_j| \leq t, 
\end{equation}where where $\hat{\bm w}=|\hat{\bm \beta}_{Enet}|^{-\alpha}$.

Kayanan and Wijekoon (2020) proposed the generalized version of LARS (GLARS) algorithm that combines LASSO and biased estimators such as RE, AURE, LE, AULE, PCRE,r-k class estimator and r-d class estimator. Further, they have shown that the combination of LASSO and r-d class estimator performed well in the high dimensional linear regression model when high multicollinearity exits among the predictor variables.

In this article, we propose improved version of GLARS algorithm that can be combined adaptive LASSO with other biased estimators such as AURE, LE, AULE, PCRE, r-k class and r-d class estimators. Further, we compared the prediction performance of the proposed algorithm with existing algorithms of adaptive LASSO and adaptive ENet using a Monte-Carlo simulation study and a real-world example.

The rest of the article is organized as follows: Section 2 presents the proposed adaptive GLARS algorithm, Section 3 evaluates the performance of the proposed algorithm, and Section 4 concludes the article.

\section{Adaptive GLARS algorithm for LASSO}\label{sec2}
Based on the methodology outlined by Kayanan and Wijekoon (2020), we propose the adaptive GLARS algorithm as follows:
\begin{algorithm}[H]
	\caption{Adaptive GLARS}\label{ch6al2}
	\begin{algorithmic}[1]
	
\State Standardize the predictor variables $\bm X$ to have a mean of zero and a standard deviation of one, and the response variable $\bm y$ to have a mean of zero. 
\State Define $\hat{\bm w}=\hat{|\bm \beta}_{G}|^{-\alpha}$ for $\alpha>0$, where $\hat{\bm \beta}_{G}$ is the general form of the biased estimators defined in equation (\ref{gef}), and $\bm X=\dfrac{\bm X}{\hat{\bm w}}$.
\State Initialize the estimated value of $\bm \beta$ as $\hat{\bm \beta} = 0$, and set the residual $ \bm r_0 = \bm y$.
\State Identify the predictor variable most correlated with $\bm r_0$ by:\begin{itemize}
    \item Calculate $X_{j1} = \max_j |Cor(X_j, \bm r_0|$ for $j = 1, 2, ..., p$.
\item Increase the estimate of $\hat{\beta}_{j1}$ from 0 until another predictor $X_{j2}$ has a high correlation with the current residual as $X_{j1}$ does.
\item Proceed in the equiangular direction between $X_{j1}$ and $X_{j2}$.
\item Similarly, each subsequent variable $X_{ji}$ earns its way into the active set, and proceed in the equiangular direction between all selected predictors
\item Update coefficient estimates using the formula:
\begin{equation}
    \hat{\bm\beta}_{ji}=\hat{\bm\beta}_{j(i-1)}+\rho_i \bm u_i,
\end{equation}where $\alpha_i$ is a value between 0 and 1 representing the distance the estimate moves before another variable enters the model, and $\bm u_i$ is the equiangular vector.
\item Calculate the direction $\bm u_i$ using:
\begin{equation}
   \bm u_i=\bm G_E(\bm E'_i \bm X'\bm X\bm E_i )^{-1} \bm E'_i \bm X'\bm r_{i-1},
\end{equation}where $\bf Ei$ is the matrix with columns $(e_{j1}, e_{j2}, ..., e_{ji})$, $e_j$ is the $j$-th standard unit vector in $\mathbb{R}^p$ with the indices of selected variables, and $\bm G_E$ depends on the specific estimator which can be substituted by respective expressions for any of estimators of our interest as listed in Table \ref{tab:l1}.
\item Update $\rho_i$ as: 
\begin{equation}
    \rho_i=\min\left\{\rho^+_{ji},\;\rho^-_{ji},\;\rho^*_{ji}\right\}\in[0,1]
\end{equation}where\begin{equation}
    \rho^\pm_{ji}=\frac{Cor(\bm r_{i-1},X_{ji})\pm Cor(\bm r_{i-1},X_{j})}{Cor(\bm r_{i-1},X_{ji})\pm Cor(\bm X\bm u_i,X_{j})} \text{for any }j\text{ such that }\hat{\bm\beta}_{j(i-1)}=0,
\end{equation}and\begin{equation}
    \rho^*_{ji}=-\frac{\hat{\bm\beta}_{j(i-1)}}{\bm u_i}\text{ for any }j\text{ such that }\hat{\bm\beta}_{j(i-1)}\neq 0.
\end{equation}  
\item If $\rho_i=\rho^*_{ji}$, update $\bm E_i$ by removing the column $e_j$ from $\bm E_{i-1}$. Calculate the new residual $\bm r_i$ as:\begin{equation}
    \bm r_i=\bm r_{i-1}-\rho_i\bm X\bm u_i,
\end{equation}and move to the next step where $j_{i+1}$ is the value of $j$ such that $ \rho_i=\rho^+_{ji}$ or $ \rho_i=\rho^-_{ji}$ or $\rho^*_{ji}$.
\item End this step when $\rho_i=1$.
\end{itemize}
\State Output $\hat{\bm\beta}_{adp}=\dfrac{\hat{\bm \beta}}{\hat{\bm w}}$.
\end{algorithmic}
\end{algorithm}

\begin{table}[H]
\footnotesize
\captionsetup{font=footnotesize}
\centering
\caption{$\bm G_E$ of the estimators for GLARS}
\label{tab:l1}       
\begin{tabular}{l|l}
\hline\noalign{\smallskip}
Estimators & $\bm G_E$   \\
\noalign{\smallskip}\hline\noalign{\smallskip}
OLSE&$\bm E_i$\\
RE&$\bm E_i (\bm E'_i (\bm X'\bm X+k\bm I)\bm E_i )^{-1} (\bm E'_i \bm X'\bm X\bm E_i )$\\
AURE&$\bm E_i(\bm I_{p_{E}}-k^2(\bm E'_i(\bm X'\bm X+k\bm I)\bm E_i )^{-2})$\\
LE&$\bm E_i(\bm E'_i(\bm X'\bm X+\bm I)\bm E_i )^{-1} (\bm E'_i(\bm X'\bm X+d\bm I)\bm E_i )$\\
AULE&$\bm E_i(\bm I_{p_{E}}-(1-d)^2(\bm E'_i(\bm X'\bm X+\bm I)\bm E_i )^{-2})$\\
PCRE&$ \bm T_{h_E}\bm T'_{h_E} \bm E_i$\\
r-k class&$\bm T_{h_E}\bm T'_{h_E} \bm E_i (\bm E'_i (\bm X'\bm X+k\bm I)\bm E_i )^{-1}(\bm E'_i \bm X'\bm X\bm E_i )$\\
r-d class&$ \bm T_{h_E}\bm T'_{h_E} \bm E_i(\bm E'_i(\bm X'\bm X+\bm I)\bm E_i )^{-1} (\bm E'_i(\bm X'\bm X+d\bm I)\bm E_i )$\\
\noalign{\smallskip}\hline
\end{tabular} 
\end{table}In Table \ref{tab:l1}, $\bm I_{p_{E}}$ is the $p_{E}\times p_{E}$ identity matrix, $p_{E}$ is the number of  selected variables in each subsequent step, and $\bm T_{h_E}=(t_{1},t_{2},...,t_{h_E})$ is the first $h_E$ columns of the standardized eigenvectors of $\bm E'_i \bm X'\bm X\bm E_i $. 

The adaptive GLARS algorithm iteratively updates combined estimates of adaptive LASSO and other estimators. Evaluation of prediction performance relies on the Root Mean Square Error (RMSE) criterion, elaborated in Section 3. Adaptive GLARS facilitates the integration of adaptive LASSO with any estimator listed in Table \ref{tab:l1}.

It is worth noting that when $\bm G_E$ corresponds to the expressions of OLSE and RE, adaptive GLARS provides solutions akin to adaptive LASSO and adaptive ENet, respectively. For ease of reference, we denote adaptive GLARS as adpLARS-LASSO, adpLARS-EN, adpLARS-AURE, adpLARS-LE, adpLARS-AULE, adpLARS-PCRE, adpLARS-rk, and adpLARS-rd when $\bm G_E$ corresponds to the expressions of OLSE, RE, AURE, LE, AULE, PCRE, r-k class, and r-d class estimators, respectively.


We can use two-dimensional cross-validation to find the suitable value of $\alpha$ and shrinkage parameter $k$ or $d$ for adaptive GLARS.

\section{Performance of the adaptive GLARS algorithms}\label{sec3}
Proposed algorithms are compared using the RMSE criterion, which is the expected prediction error of the algorithms, and is defined as\begin{equation}
RMSE(\hat{\bm \beta})=\sqrt{\frac{1}{n}({\bm y_{new}}-\bm X_{new}\hat{\bm \beta})'({\bm y_{new}}-\bm X_{new}\hat{\bm \beta})}
\end{equation}
where $(\bm y_{new},\bm X_{new})$ denotes the new data which are not used to obtain the parameter estimates, and $\hat{\bm \beta}$ is the estimated value of $\bm \beta$ using the respective algorithm. A Monte Carlo simulation study and the real-world examples are used for the comparison. 

\subsection{Simulation study}
According to McDonald and Galarneau (1975), first we generate the predictor variables by using the following formula:
\begin{equation}
x_{i,j}=\sqrt{(1-\rho^{2})} z_{i,j}+\rho z_{i,m+1}  \qquad;i=1,2,...,n.\;\;  j=1,2,...,m.
\end{equation}
where $z_{i,j}$ is an independent standard normal pseudo random number, and $\rho$ is the theoretical correlation between any two explanatory variables. 

In this study, we have used a linear regression model of 100 observations and 20 predictors.
 A dependent variable is generated by using the following equation
\begin{equation}
y_{i}=\beta_{1} x_{i,1}+\beta_{2} x_{i,2}+...+\beta_{5} x_{i,20}+\epsilon_{i}  \qquad;i=1,2,...,100.
\end{equation}
where $\epsilon_{i}$ is a normal pseudo random number with mean zero and common variance $\sigma^2$.
We choose $\bm \beta=(\beta_{1}, \beta_{2} , ...,\beta_{20})$ as the normalized eigenvector corresponding to the largest eigenvalue of $\bm X'\bm X$ for which $\bm \beta'\bm\beta=1$. To investigate the effects of different degrees of multicollinearity on the estimators, we choose $\rho=(0.5,\; 0.7, \;0.9)$, which represents weak, moderated and high multicollinearity. For the analysis, we have simulated 50 data sets consisting of 50 observations to fit the model and 50 observations to calculate the RMSE.

\begin{figure}[H]
\centering
    \includegraphics[width=0.57\textwidth]{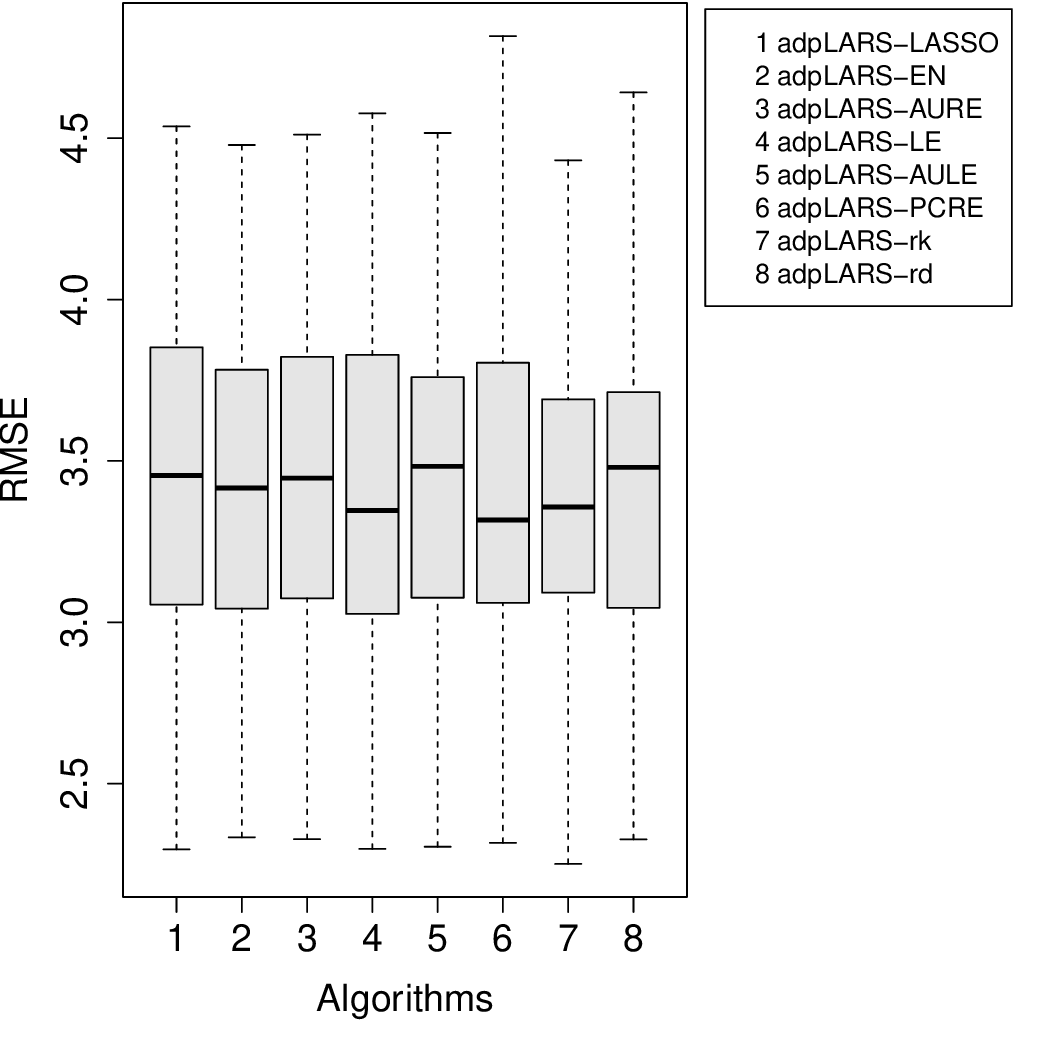}
\caption{     Cross-validated RMSE values of the adaptive GLARS algorithms when $\rho=0.5$. }
\label{ch6f5}       
\end{figure}

\begin{figure}[H]
\centering
    \includegraphics[width=0.57\textwidth]{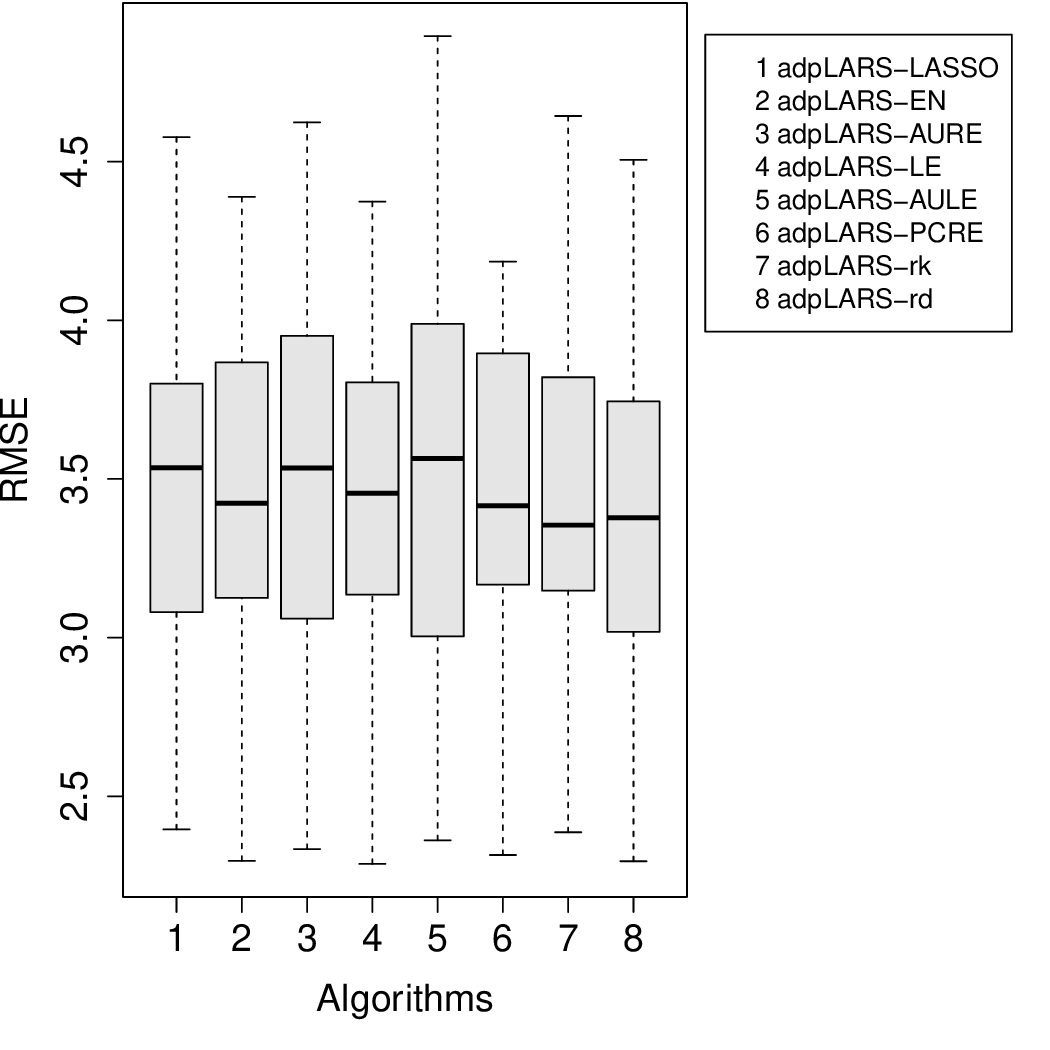}
\caption{     Cross-validated RMSE values of the adaptive GLARS algorithms when $\rho=0.7$. }
\label{ch6f6}       
\end{figure}

\begin{figure}[H]
\centering
    \includegraphics[width=0.57\textwidth]{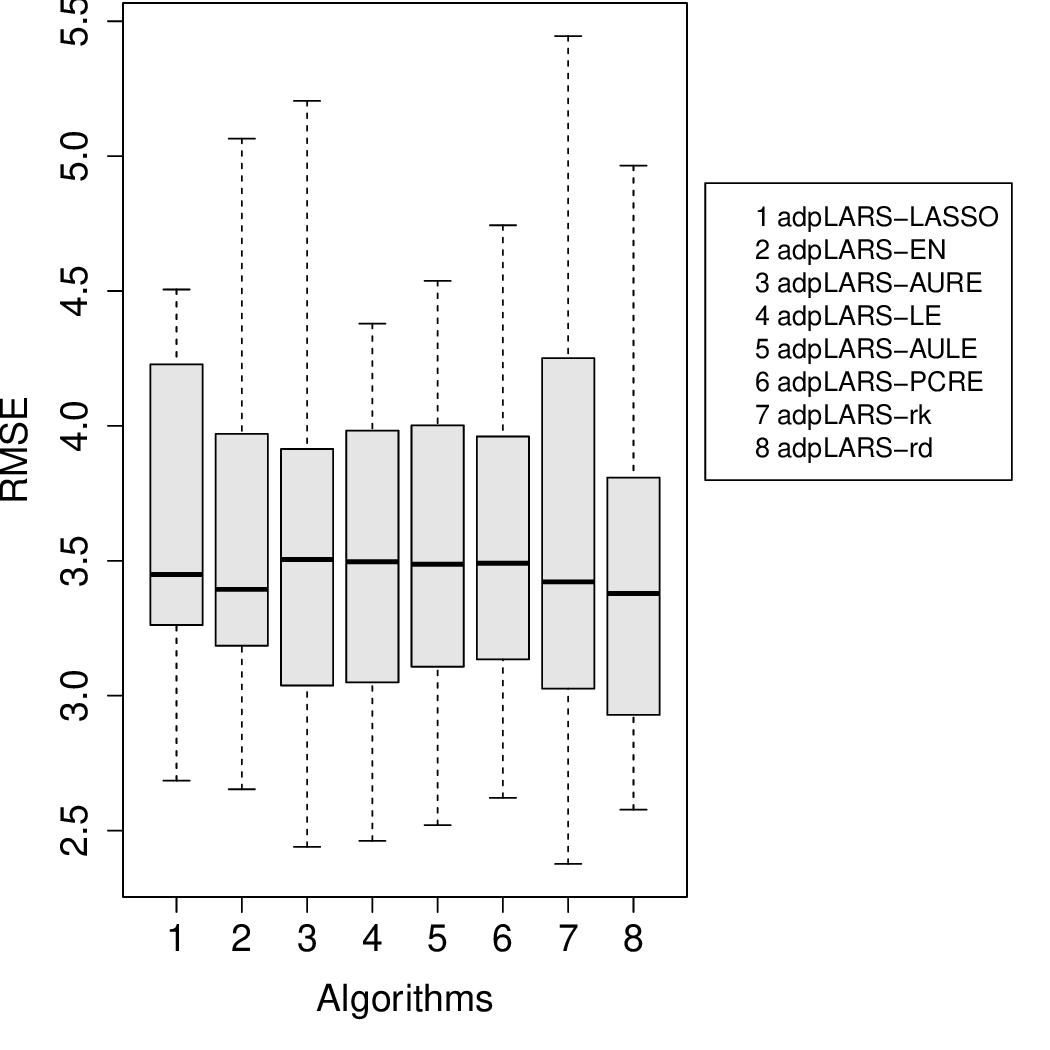}
\caption{     Cross-validated RMSE values of the adaptive GLARS algorithms when $\rho=0.9$. }
\label{ch6f7}       
\end{figure}

The Cross-validated RMSE of the adaptive GLARS algorithms are displayed in Fig. \ref{ch6f5} - Fig. \ref{ch6f7}, and the median cross-validated RMSE of the algorithms are displayed in Table \ref{ch6t6} - Table \ref{ch6t8}. 

\begin{table}[H]
\centering
\caption{     Median Cross-validated RMSE values of the adaptive GLARS algorithms when $\rho=0.5$.}
\label{ch6t6}       
\begin{tabular}{lccccc}
\hline\noalign{\smallskip}
Algorithms & RMSE & ($k,\;d$)&$\alpha$&$t$&Selected variables   \\
\noalign{\smallskip}\hline\noalign{\smallskip}
adpLARS-LASSO&3.45489&	--&1&	6.6635	&16\\
adpLARS-EN& 3.41614&	0.2&1&	7.5795&	17\\
adpLARS-AURE&3.44668&	1.0&1&	7.1685&	17\\
adpLARS-LE&3.34648&	0.3&	1& 7.1018&	15\\
adpLARS-AULE&3.48312&	0.2&1&	8.0718&	16\\
adpLARS-PCRE&\textbf{3.31719}&--&1&		6.5019&	16\\
adpLARS-rk&3.35712&	0.2&	1&6.0726&	17\\
adpLARS-rd& 3.47994&	0.99&1&	6.5019&	16\\
\noalign{\smallskip}\hline
\end{tabular}
\end{table}

\begin{table}[H]
\centering
\caption{     Median Cross-validated RMSE values of the adaptive GLARS algorithms when $\rho=0.7$.}
\label{ch6t7}       
\begin{tabular}{lccccc}
\hline\noalign{\smallskip}
Algorithms & RMSE & ($k,\;d$)&$\alpha$&$t$&Selected variables   \\
\noalign{\smallskip}\hline\noalign{\smallskip}
adpLARS-LASSO&3.53553&	--&1&	 8.7067	&16\\
adpLARS-EN& 3.42320&	0.3&0.5&	8.9330&	17\\
adpLARS-AURE&3.53440&	0.7&1&	8.0610&	17\\
adpLARS-LE&3.45469&	0.1&	0.5& 9.1520&	15\\
adpLARS-AULE&3.56472&	0.1&1&	8.0821&	16\\
adpLARS-PCRE&3.41530&--&1&		9.5873&	16\\
adpLARS-rk&\textbf{3.35452}&	0.1&	1&8.9412&	16\\
adpLARS-rd& 3.37755&	0.2&1&	8.8207&	16\\
\noalign{\smallskip}\hline
\end{tabular}
\end{table}

\begin{table}[H]
\centering
\caption{     Median Cross-validated RMSE values of the adaptive GLARS algorithms when $\rho=0.9$.}
\label{ch6t8}       
\begin{tabular}{lccccc}
\hline\noalign{\smallskip}
Algorithms & RMSE & ($k,\;d$)&$\alpha$&$t$&Selected variables   \\
\noalign{\smallskip}\hline\noalign{\smallskip}
adpLARS-LASSO&3.44950&	--&0.5&	 4.0460	&15\\
adpLARS-EN& 3.39404&	1.0&1&	8.2710&	17\\
adpLARS-AURE&3.50448&	0.9&1&	10.045&	17\\
adpLARS-LE&3.49651&	0.1&	0.5& 10.005&	15\\
adpLARS-AULE&3.48735&	0.1&0.5&	8.0684&	16\\
adpLARS-PCRE& 3.49078&--&0.5&		10.682&	17\\
adpLARS-rk&3.42176&	0.3&	1&10.433&	16\\
adpLARS-rd& \textbf{3.37842}&	0.99&0.5&	7.0576&	15\\
\noalign{\smallskip}\hline
\end{tabular}
\end{table}

Based on the insights gathered from Fig. \ref{ch6f5} to Fig. \ref{ch6f7} and Table \ref{ch6t6} to Table \ref{ch6t8}, it is evident that the adpLARS-PCRE, adpLARS-rk, and adpLARS-rd algorithms consistently demonstrate superior performance in terms of RMSE criterion compared to other adaptive GLARS algorithms across varying degrees of multicollinearity, from weak to moderate and high levels, respectively.

\subsection{Real-world example}

In our analysis, we utilized the Prostate Cancer Data (Stamey et al., 1989), a well-established dataset explored by  Tibshirani (1996), Efron et al. (2004), and Zou and Hastie (2005) to evaluate the efficacy of LASSO, LARS algorithm, and Enet.

The Prostate Cancer Data comprises eight clinical metrics: log cancer volume (lcavol), log prostate weight (lweight), age, log of benign prostatic hyperplasia volume (lbph), seminal vesicle invasion (svi), log capsular penetration (lcp), Gleason score (gleason), and percentage Gleason score 4 or 5 (pgg45). The response variable is the log of prostate-specific antigen (lpsa), with a dataset size of 97 observations. Notably, the predictor variables exhibit Variance Inflation Factor (VIF) values of 3.09, 2.97, 2.47, 2.05, 1.95, 1.37, 1.36, and 1.32, indicating considerable multicollinearity, as evidenced by a high condition number of 243. This dataset is readily available within the "lasso2" R package. Our analysis involved fitting the model with 67 observations and computing the Root Mean Square Error (RMSE) using 30 observations.

\begin{table}[H]
\centering
\caption{     Cross-validated RMSE values of Prostate Cancer Data using adaptive GLARS.}
\label{ch6t9}       
\begin{tabular}{lccccc}
\hline\noalign{\smallskip}
Algorithms & RMSE & ($k,\;d$)&$\alpha$&$t$&Selected variables   \\
\noalign{\smallskip}\hline\noalign{\smallskip}
adpLARS-LASSO&0.77653&	--&	0.2&1.57112&	7\\
adpLARS-EN&0.78716	&0.3&1&	0.80638&	7\\
adpLARS-AURE&0.80638&	1.0&0.9&	0.80638&	7\\
adpLARS-LE&0.80014&	0.1&0.5&	1.45884&	7\\
adpLARS-AULE&0.79046&	0.2&1&	1.31322&	6\\
adpLARS-PCRE&0.76890	&--&0.9&	1.44929&	7\\
adpLARS-rk&0.77698&	0.2&0.9&	1.36273&	7\\
adpLARS-rd&\textbf{0.76854}&	0.7&0.9&	1.44764&	7\\
\noalign{\smallskip}\hline
\end{tabular}
\end{table}

The cross-validated RMSE values obtained through the adaptive GLARS algorithms are summarized in Table \ref{ch6t9}. Upon examining Table \ref{ch6t9}, it becomes evident that the adpLARS-rd algorithm outperforms other algorithms when applied to the Prostate Cancer Data.

\section{Conclusions}\label{sec4}

In conclusion, this study clearly shows that the adpLARS-rk and adpLARS-rd algorithms work well for dealing with high dimensional linear regression problems, especially when there are many closely related independent variables. These algorithms emerge as reliable tools for tackling high dimensional regression models and offer promising avenues for future research and practical application in data-driven environments. 

\section*{Acknowledgement}
The authors are grateful to the anonymous referee for a careful
checking of the details and for helpful comments that improved this
paper.




  \begin{reference}
\item[] Efron B, Hastie T, Johnstone I and Tibshirani R (2004). Least angle regression, The Annals of statistics, 32(2), 407–499.

\item[] Kayanan M and Wijekoon P (2017). Performance of existing biased estimators and the respective predictors in a misspecified linear regression model, Open Journal of Statistics, 7(5), 876–900.

\item[] Kayanan M and Wijekoon P (2020). Variable selection via biased estimators in the linear regression model, Open Journal of Statistics, 10, 113–126.

\item[] McDonald GC and Galarneau DI (1975). A monte carlo evaluation of some ridge-type estimators, Journal of the American Statistical Association, 70(350), 407–416.

\item[] Stamey TA, Kabalin JN, et al. (1989). Prostate specific antigen in the diagnosis and treatment of adenocarcinoma of the prostate: Ii. radical prostatectomy treated patients, Journal of Urology, 141(5), 1076–1083.

\item[] Tibshirani R (1996). Regression shrinkage and selection via the lasso, Journal of the Royal Statistical Society: Series B (Methodological), 58(1), 267–288.

\item[] Zou H (2006). The adaptive lasso and its oracle properties, Journal of the American Statistical Association, 101(476), 1418–1429.

\item[] Zou H and Hastie T (2005). Regularization and variable selection via the elastic net, Journal of the Royal Statistical Society: Series B, 67(2), 301–320.

\item[] Zou H and Zhang HH (2009). On the adaptive elastic-net with a diverging number of parameters, The Annals of Statistics, 37(4), 1733–1751
\end{reference}
\end{document}